\pdfoutput=1

\documentclass{appolb}

\usepackage{amssymb}
\usepackage{hyperref}
\usepackage{graphicx}
\usepackage{amsmath}
\usepackage[numbers,sectionbib,sort&compress]{natbib}

\begin{document}

\title{Hollowness in $pp$ scattering at the LHC%
\thanks{Talk presented by WB at Excited QCD 2017, 7-13 May 2017, Sintra, Portugal}
\thanks{Supported by Polish National Science Center grant 2015/19/B/ST2/00937, by Spanish Mineco Grant FIS2014-59386-P, 
and by Junta de Andaluc\'{\i}a grant FQM225-05.}}

\author{Wojciech Broniowski$^{1,2}$\thanks{Wojciech.Broniowski@ifj.edu.pl} and Enrique Ruiz Arriola$^3$\thanks{earriola@ugr.es}
\address{${}^{1}$The H. Niewodnicza\'nski Institute of Nuclear Physics, \\ Polish Academy of Sciences, PL-31342~Cracow, Poland}
\address{${}^{2}$Institute of Physics, Jan Kochanowski University, PL-25406~Kielce, Poland}
\address{${}^{3}$Departamento de F\'isica At\'omica, Molecular y Nuclear and \\ Instituto Carlos I de Fisica Te\'orica y Computacional,  Universidad de Granada, E-18071 Granada, Spain}}

\maketitle

\begin{abstract}
We examine how the effect of hollowness in $pp$ scattering at the LHC 
(minimum of the inelasticity profile at zero impact parameter) depends on modeling of the 
phase of the elastic scattering amplitude as a function of the momentum transfer. 
We study the cases of the constant phase, the Bailly, and  the so called 
standard parameterizations. It is found that the 2D hollowness holds in the first two cases, 
whereas the 3D hollowness is a robust effect, holding for all explored cases. 
\end{abstract}


In this contribution we focus on the aspects of the alleged {\em
hollowness} effect in $pp$ scattering not covered in our previous
paper~\cite{RuizArriola:2016ihz} and
talks~\cite{Arriola:2016bxa,Broniowski:2017aaf}, where the basic
concepts and further details of the presented analysis may be found.
The recent TOTEM~\cite{Antchev:2013gaa} and ATLAS
(ALFA)~\cite{Aad:2014dca} data for the differential elastic cross
section for $pp$ collisions at $\sqrt{s}=7$~TeV and
$\sqrt{s}=8$~TeV~\cite{Antchev:2013paa,Aaboud:2016ijx} suggest a
stunning behavior (impossible to explain on classical grounds), where
more inelasticity in the reaction occurs when the protons collide at
an impact parameter $b$ of a fraction of a fermi, than for head-on
collisions.
Here we discuss the sensitivity of this hollowness feature on modeling
of the phase of the elastic scattering amplitude as a function of the
momentum transfer. In previous analyses~\cite{RuizArriola:2016ihz,Alkin:2014rfa,Dremin:2014eva,Dremin:2014spa,Dremin:2016ugi,
Dremin:2017ylm,Dremin:2017qfo,Anisovich:2014wha,Albacete:2016pmp,Troshin:2016opy,Troshin:2016frs,Troshin:2017zmg,Troshin:2017ucy} this effect was not treated
with sufficient attention.

\begin{figure}
\begin{center}
\includegraphics[width=0.48\textwidth]{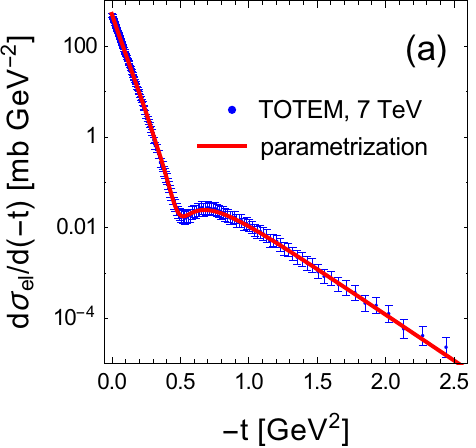} \hfill  \includegraphics[width=0.46\textwidth]{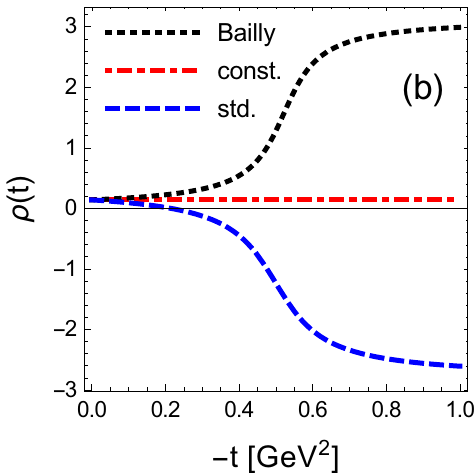} 
\end{center}
\vspace{-3mm}
\caption{(a) Data for the differential elastic strong-interaction cross section at the LHC energy 
of  $\sqrt{s}=7$~TeV~\cite{Antchev:2013gaa} with the overlaid fit of Eq.~\ref{eq:mBP2}. 
(b)~Phase of the strong-interaction elastic scattering amplitude, according 
to the three models of Eq.(\ref{eq:const}-\ref{eq:standard}). \label{fig:data}}
\end{figure}

In the present work we parametrize separately the absolute value and the phase of the strong elastic $pp$ scattering amplitude. For the absolute value we apply 
the form of Ref.~\cite{Fagundes:2013aja}:
\begin{eqnarray}
|{f} (s,t)| = p \left | \frac{i \sqrt{A} e^{\frac{B t}{2}}}{\left(1-\frac{t}{t_0}\right)^4}+i \sqrt{C} e^{\frac{D t}{2}+i \phi } \right | \, ,   \label{eq:mBP2}
\end{eqnarray}
where $p$ is the CM momentum, and $A$, $B$, $C$, $D$, and $t_0$ were
adjusted to the data. We neglect spin effects, hence the amplitude is
to be understood as spin-averaged.  The quality of the fit to
differential elastic cross section from the LHC data at
$\sqrt{s}=7$~TeV can be assessed from Fig.~\ref{fig:data}(a). This fit
is sensitive only to the square of the absolute value of the
amplitude, and not to its {\it phase}.  However, this is not true of
other features of $pp$ scattering, which do depend of the phase.

The $\rho(s,t)$ function is defined as the ratio of the real to imaginary parts of ${f} (s,t)$:
\begin{eqnarray}
\rho(s,t) = \frac{{\rm Re}{f} (s,t) }{{\rm Im}{f} (s,t)}, \;\;\;\;\;\;\; {f} (s,t) = \frac{i+\rho(s,t)}{\sqrt{1+\rho(s,t)^2}}|{f} (s,t)|.
\end{eqnarray}
At $t=0$, $\rho(s,0)$ can be determined when the total cross section
$\sigma_{\rm tot}(s)$ and the differential cross section extrapolated
to $t=0$ are know. In actual analyses, interference with the Coulomb
amplitude is used to determine $\rho$ (see in particular
Ref.~\cite{Antchev:2016vpy} for further information and
literature). The value of the phase at $t=0$ for $\sqrt{s}=7$~TeV has
been determined to be $\rho(7~{\rm
TeV},0)=0.145(100)$~\cite{Antchev:2013gaa}. However, one should bear
in mind that the extraction of the dependence of $\rho(s,t)$ on $t$ via
the separation of the electromagnetic and strong
amplitudes~\cite{West:1968du} is sensitive to the internal
electromagnetic structure of the proton and is subject to on-going
debate~\cite{Prochazka:2016wno}.

In this contribution we explore three popular parameterizations:
constant, 
\begin{eqnarray}
\rho(t)=\rho_0={\rm const.}, \label{eq:const}
\end{eqnarray}
with $\rho_0=0.14$, 
the Bailly et al.~\cite{Bailly:1987ki} parametrization,
\begin{eqnarray}
\rho(t)=\frac{\rho_0(s)}{1-t/t_d}, \label{eq:Bailly}
\end{eqnarray}
where $t_d=0.52~{\rm GeV}^2$ is the position of the diffractive minimum,
and the so called standard parametrization\footnote{A similar form to the standard parametrization arises in the Pomeron exchange models, see, e.g.,~\cite{Godizov:2017ygm}.}
\begin{eqnarray}
 \rho(t)=\rho _0+\frac{\left(\rho _0^2+1\right) \tau t }{\tau ^2 +t_0^2 -  \left(\rho _0 \tau +t_0\right) t}  \label{eq:standard}
\end{eqnarray}
with $t_0=0.5~{\rm GeV}^2$ and $\tau=0.1~{\rm GeV}^2$.

The $b$ representation
the scattering amplitude is defined via the
Fourier-Bessel transform of $f(s,t)$, as given by the data parametrization,
\begin{eqnarray}
2ph(b,s)= 2 \int_0^\infty q dq J_0(bq) f(s,-q^2) ={i} \left [ 1-e^{i \chi(b)}  \right ], \label{eq:eiko}
\nonumber \label{eq:invf}
\end{eqnarray} 
where we have also introduced the {\em eikonal phase} $\chi(b)$.
\begin{figure}
\begin{center}
\includegraphics[width=0.47\textwidth]{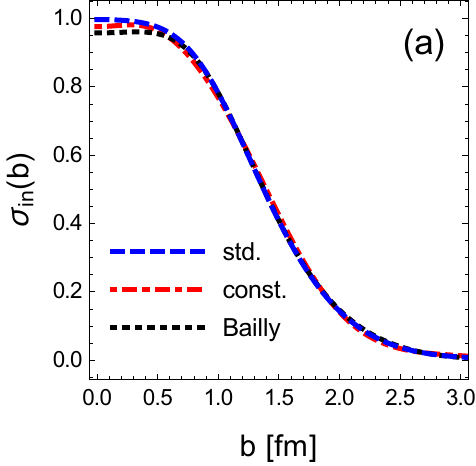} \hfill \includegraphics[width=0.48\textwidth]{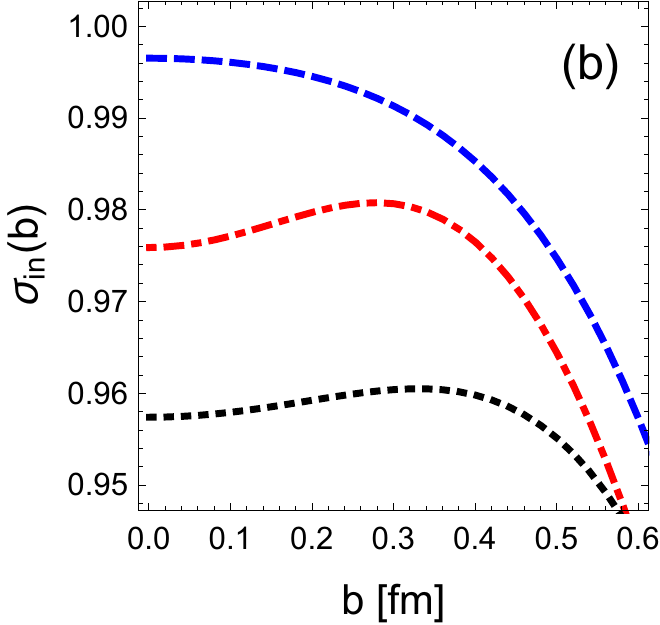}
\end{center}
\vspace{-3mm}
\caption{(a) Inelastic cross section in the impact-parameter representation for three models of $\rho(t)$ from Eq.(\ref{eq:const}-\ref{eq:standard}). 
(b)~Close-up for small values of $b$. \label{fig:sigmas}}
\end{figure} 
\begin{figure}
\begin{center}
\includegraphics[width=0.47\textwidth]{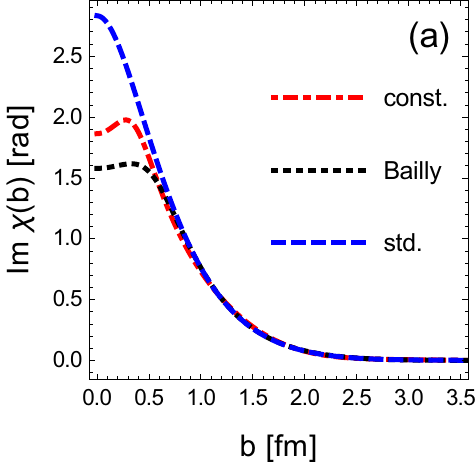} \hfill \includegraphics[width=0.48\textwidth]{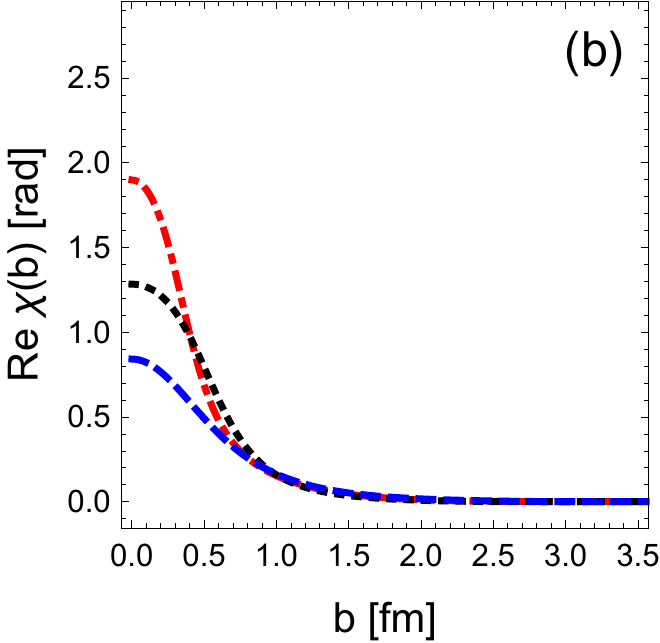}
\end{center}
\vspace{-3mm}
\caption{Same as in Fig.~\ref{fig:sigmas}, but for the imaginary (a) and real (b) parts of the eikonal scattering phase. \label{fig:eiko}}
\end{figure} 
The equation for the inelastic cross section is
\begin{eqnarray} 
\sigma_{\rm in} &\equiv& \sigma_T - \sigma_{\rm el} = \int d^2 b \left( 4p {\rm Im} h(b,s) -  4p^2|h(b,s)|^2 \right ) ,  \label{eq:sin}
\end{eqnarray} 
where the integrand is the {\em inelasticity profile}, with $0\le \sigma_{\rm in}(b) \le 1$.

\begin{figure}
\begin{center}
\includegraphics[width=0.47\textwidth]{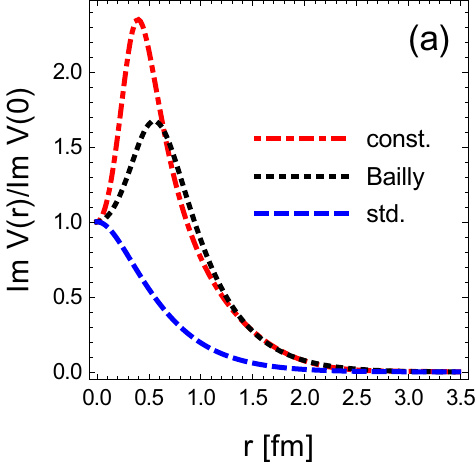} \hfill \includegraphics[width=0.47\textwidth]{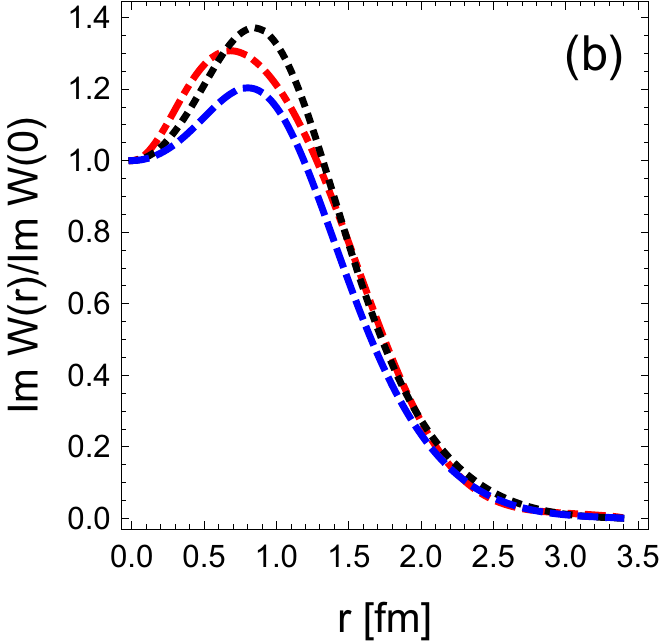}
\end{center}
\vspace{-3mm}
\caption{Imaginary parts of the optical potential $V(r)$ (a) and the on-shell optical 
potential $W(r)$ (b), introduced in Refs.~\cite{Arriola:2016bxa,RuizArriola:2016ihz}, plotted for parameterizations of $\rho(t)$ 
from Eq.(\ref{eq:const}-\ref{eq:standard}). \label{fig:uw}}
\end{figure}  
 
In Fig.~\ref{fig:sigmas} we show $\sigma_{\rm tot} (b)$ for three
parameterizations $\rho(t)$ of Eq.(\ref{eq:const}-\ref{eq:standard}).
We note that hollowness appears for the first two models, whereas it
is absent for the ``standard'' parametrization.  The imaginary and
real parts of the eikonal phase are presented in Fig.~\ref{fig:eiko},
where we note the corresponding dips at $b=0$ for the imaginary parts
-- a feature that follows from the eikonal
formalism~\cite{Broniowski:2017aaf}.

Finally, in Fig.~\ref{fig:uw} we show the imaginary parts of the optical potential $V(r)$ and the on-shell optical 
potential $W(r)$, introduced in Refs.~\cite{Arriola:2016bxa,RuizArriola:2016ihz}. We note that in this 3D picture of $pp$
scattering, hollowness occurs for all the considered models of $\rho(t)$.

To summarize, a firm establishment of the 2D hollowness requires a
careful determination of the phase of the strong-interaction elastic
amplitude. On the other hand, hollowness in 3D is a robust effect. The
intriguing property of hollowness must have quantum
origin~\cite{Arriola:2016bxa,Broniowski:2017aaf}, hence touches upon
very basic features of the scattering mechanism. Hopefully, future
data and more refined analyses based on the Coulomb separation will
sort out the issue in 2D.

\bibliography{NN-high-energy}
 
\end{document}